\documentclass[preprint,12pt]{elsarticle}

\usepackage{amssymb}
\usepackage{algpseudocode}
\usepackage{algorithm}
\usepackage{bm}
\usepackage{listings}

\usepackage{lineno}
\usepackage{pgfplots}
\usepackage{pstricks,pst-node, pst-fun}
\usepackage{dsfont}
\usepackage{amssymb}
\usepackage[british,UKenglish]{babel}
\usepackage[utf8]{inputenc}
\usepackage[a4paper,left=3cm,right=2cm,top=2.5cm,bottom=2.5cm]{geometry}
\usepackage{amsmath}
\usepackage{relsize}
\usepackage{setspace}
\usepackage{subfig}
\DeclareMathAlphabet{\mathpzc}{OT1}{pzc}{m}{it}

\DeclareFixedFont{\ttb}{T1}{txtt}{bx}{n}{12} 
\DeclareFixedFont{\ttm}{T1}{txtt}{m}{n}{12}  

\usepackage{color}
\definecolor{deepblue}{rgb}{0,0,0.5}
\definecolor{deepred}{rgb}{0.6,0,0}
\definecolor{deepgreen}{rgb}{0,0.5,0}

\usepackage{listings}

\newcommand\pythonstyle{\lstset{
language=Python,
basicstyle=\ttm,
morekeywords={self,False,True},              
keywordstyle=\ttb\color{deepblue},
emph={MyClass,__init__},          
emphstyle=\ttb\color{deepred},    
stringstyle=\color{deepgreen},
frame=tb,                         
showstringspaces=false
}}

\lstnewenvironment{python}[1][]
{
\pythonstyle
\lstset{#1}
}
{}

\newcommand\pythoninline[1]{{\pythonstyle\lstinline!#1!}}

\journal{Journal Name}

\begin{document}

\begin{frontmatter}






\title{Probabilistic combination of loads in topology optimization designs via cumulative damage criteria}


\author{Luis Irastorza-Valera$^{a}$, Luis Saucedo-Mora$^{a}$}

\address{$^a$ E.T.S. de Ingeniería Aeronáutica y del Espacio, Universidad Politécnica de Madrid, Pza. Cardenal Cisneros 3, 28040, Madrid, Spain}
\vspace{0.3cm}

\begin{abstract}
 
Topology optimization (TO) is a well-established methodology for structural design under user-defined constraints, e.g. minimum volume and maximum stiffness. However, such methods have traditionally been applied to static, deterministic loading, in which modulus, position and direction are known and invariant. This is against the probabilistic load combination used in the structural engineering designs, and entails two important shortcomings. 

The first one is related to maintenance and reliability: static loading fails to consider naturally occurring uncertainties in the loading process, measurements or regular service; also ignoring (quantitatively) unforeseen phenomena such as vibrations, and the material's behavior is assumed linear isotropic, ignoring fatigue, plasticity and anisotropy in functionally-graded materials.
The second one concerns optimality itself: often times, the structure presented as "optimal" in fact over-estimates loading and thus wastes material by exceeding its real needs and/or distributing it poorly throughout the design dominion.

In this article, a probabilistic framework is presented: uncertain and pseudo-dynamic loading is introduced to create robust topologies via a reinforced SIMP scheme with embedded penalization addressing fatigue damage, layer direction, mechanical response (traction/compression) and yield limit (von Mises equivalent stress). This computationally efficient framework is applied to various loading scenarios, generating diverse designs for the same volume fraction constraint with improved and more realistic performances. Under the proposed method, if loads are permanent and damage isotropic, the methodology converges to the traditional (deterministic) topological optimization results. Future ramifications of this work are pondered, especially regarding metamaterial design. 
\end{abstract}


\end{frontmatter}


\section{Introduction}

Traditionally, structural design in industry has sought a trade-off between mechanical (e.g. maximum stiffness) and economic goals (e.g. minimum material). Frames \cite{Michell1904}, flexural meshes \cite{Rozvany1972,Prager1977} and the continuum \cite{Bendse1988,Bendse1989} have been successfully tackled in that regard by Topology Optimization (TO), in which the aim is changing the genus (i.e., number of holes) in the prototype to fit the designer's purposes. 

Solid Isotropic Material with Penalization (SIMP) \cite{Bendse1989,Rozvany2000} is the most used TO technique, where material is distributed along a given spatial dominion $\Omega \in \mathbb{R}^n$ while minimizing elastic strain energy (half the compliance, $\mathbf{c} = \mathbf{u^T Ku}$) within a required volume fraction $f=V_f/V_0$. This is done by penalizing the stiffness (Young's modulus) of under-loaded elements at each iteration by an exponent $p$, as in $E^t=E_0\rho^p$, $\rho$ being the material's density (from 0 - void - to 1 - full -). SIMP is popular due to its conceptual simplicity, easily implementable on code in two \cite{Sigmund2001,Andreassen2010} and three dimensions \cite{Liu2014}. 


The TO framework was primarily conceived for static settings, i.e. time-invariant boundary conditions and loading. This is the case for most boundary value problems (BVP). However, stationary conditions are quite reductionist, considering real performance entails naturally occurring uncertainties disrupting this determinism (load misplacement by vibrations, faulty measurements, slight variations in modulus and direction, etc.) and wear/tear (decay in material properties, potentially leading to fracture). Loading uncertainty has been analyzed by several authors using different mathematical tools (statistics, model order reduction, etc.) \cite{Dunning2011,Zhao2014,Torres2021}. 

Structural design applies the superposition principle to loading: the individual loads can be added on top of each other so that the resulting topology endures their combined effects, often linearly, as a truss-like approximation of the continuum \cite{Rozvany2014,Cui2019,Shi2023}, or perhaps by integrating force fields (pressure) \cite{Bourdin2003}. 

In TO, this is usually done by summing up the full compliance contributions of each individual load \cite{Sigmund2001,Andreassen2010,Liu2014}. While safe, the "load envelope" approach is demonstrably sub-optimal (excessively conservative) in its material distribution, and perhaps counter-productive under certain loading conditions. 

Statistics can convey the uncertainty derived from unforeseen non-static eventualities (vibration, impacts) during a structure's life cycle in the TO process, providing realistic sizing based on their average frequency of appearance \cite{Dunning2011}. This gave rise to reliability/performance-based topology optimization (RBTO/PBTO) and inverse optimum safety factors (IOSF) \cite{Kharmanda2019,Kharmanda2020}, striving for a more robust, probabilistic approach and more suitable for additive manufacturing \cite{KaramoozRavari2014}. This is coherent with practical implementations in  industry, based on confidence intervals (e.g. six-sigma).

Considering TO is a non-convex problem \cite{Abdelhamid2021}, reaching the global minimum is not guaranteed, so gradient descent often finds local minima instead. This can be dealt with via several methods \cite{Liao2019}, such as filtering and relaxation \cite{Sigmund1998} and model order reduction (MOR) \cite{Torres2021,Yano2021}. Interestingly, TO's non-convexity leaves room for other design constraints to be introduced, like explicit material properties (e.g. the Updated Properties Model, UPM \cite{SaucedoMora2023}) and even multi-material proposals \cite{Zuo2016}, fatigue damage, manufacturing restrictions, etc. 

Some material properties besides the Young's modulus could be included as constraints (maximum stress \cite{Paris2008,Holmberg2013,DeLeon2015,Yang2018}, fatigue \cite{Holmberg2014,Oest2017,Suresh2019}, manufacturability \cite{Mirzendehdel2016}, load uncertainty \cite{Chen2020}) to obtain more realistic and robust optimized designs, reflecting actual material behavior. For instance, concrete-based materials work great under compression but poorly under traction, whereas the opposite is true for steel and many other commonly-used metals, due to buckling when the struts are slender enough. Implementation of such criteria should be straightforward and not excessively demanding, computationally speaking.


Damage should be taken into account for maintenance purposes, since it alters the material's effective mechanical properties. For the usual static case, simple stress-based formulations (Rankine, Saint-Venant, Tresca, von Mises, etc.) can be implemented, whereas fatigue damage induced by dynamic loads requires more elaborate stress (Wöller's diagram), strain (e.g., Smith-Watson-Topper's model \cite{Zhu2011}) and/or energy criteria (e.g. fracture mechanics) integrated over time. 

The latter is directly related to TO's objective (SIMP) - minimizing compliance, i.e. elastic strain energy. Perhaps total strain energy (elastic and plastic) can be used as an index for cumulative fatigue damage in itself - including explicit fracture considerations (e.g., crack growth) \cite{Golos1988,Golos1989} -  useful for damage prediction \cite{Cha2015}.

After a brief summary of the state of the art in this introductory first section, the proposed methodology will be explained in Section 2. Some illustrative results will be analyzed in Section 3 whereas Section 4 will provides some conclusions and future research lines.

\section{Methodology}
This section will delve into the tools used in this article, the proposed novel approaches and their practical implementation through code. Two  dedicated subsections will provide details about loading uncertainty and embedded damage criteria, respectively.

\subsection{Probabilistic Uncertainty}

Statistics can introduce the uncertainties derived from normal (vibrations) and “abnormal” eventualities (impacts) during a structure's life cycle in the TO process, providing realistic sizing based on their average frequency of appearance \cite{Dunning2011}. If each observed load is associated with a cumulative known damage contribution, the resulting optimized structure will be adequate in size, robust and reliable in predictive maintenance terms within a measurable confidence interval, which is contingent in industry. Uncertainty is induced via loading. 

\subsubsection{Loading Types}

Several load types $i$ can be identified, each of them with an associated probability $p$ and number of cycles $N$:
\begin{itemize}
    \item \textbf{Constant loads} are always applied and so, they just need one cycle per iteration to capture their ever-present contribution ($p_i = 1$). They represent static, invariant loads as in traditional TO. 
    \item \textbf{Alternating loads} appear in alternative iterations and/or positions, with $p_i = 1/(N_{f,i})$ (given by their frequency share) for every load $i$. For instance, two loads switching between two alternative positions each iteration will have each a 50\% probability ($p_i = 1/2$). They account for regularly-timed and predictable alternative loads, generating fatigue damage to be introduced via tailored compliance penalization. 
    \item \textbf{Statistical loads} capture the uncertainty (in position, direction and modulus) associated to the loading process, both inherent (e.g. loading/unloading) and induced (vibrations). While they appear in every iteration, they have unique associated probabilities $p_{ij}$ for each cycle $j$ and load $i$, randomly drawn from statistical distributions whose means $\mu_p$ and $\mu_m$ represent their canonical position and modulus, respectively. This probabilistic approach to loading implies a confidence interval for the results, as in industry (e.g. six-sigma), conveying variance.
    \item \textbf{Occasional loads} appear every $X$ iterations ($p_i = 1/X$). Their position and modulus are static, striving to capture the effects of unforeseen eventualities, such as impacts: significant in modulus, albeit non-frequent. They leave a noticeable mark in the TO process, which reinforces the structure to face such occasions.
\end{itemize}

By including all these loading types, the TO process is not only striving for a topologically optimized structure, but also learning how to adapt it to various scenarios, which renders the resulting topology more robust. 

\subsubsection{Penalization}

Instead of applying the same static loads in each iteration loop and computing a single global compliance $c$ to update the volume distribution, as in the original SIMP approach \cite{Bendse1988}, that global loop compliance results from the superposed sum of all individual loads $i$ (from cycles $j$) applied simultaneously in that precise iteration multiplied by their assigned probabilities. This rationale resembles cumulative fatigue damage in Miner's rule - see Equation \ref{eq:Miner}:
\begin{equation}
    \\\\\ \gamma = \sum^k_{m=1} \frac{n_m}{N_m}
    \label{eq:Miner}
\end{equation}
where $\gamma$ represents the total accumulated damage as the sum of the quotients between the number of cycles $n$ undergone by the prototype and their service life $N$, for each amplitude level $m$ (out of $k$). In the proposed approach, amplitude levels are conceptually analogous to load damage contributions, given by their moduli and frequencies. 

By considering separate cyclic contributions, two additional non-deterministic dimensions are added to loading: variability and frequency. On the one hand, individualizing cycles and loads separately allows for the selection of tailored cyclic strategies for each load, including slight variations in modulus to account for uncertainties (e.g. vibrations). On the other hand, cyclic contributions can be switched on and off on demand, effectively establishing different incidence frequencies for each load. This conveys the compliance-weighted influences of each load onto the final optimized topology. 

The algorithm “learns” which loads to expect and when and reinforces the optimizing topology accordingly, even if such loads are not constantly applied (i.e. deterministic TO). For instance, one big impact load applied every $X$ iterations makes the topology more robust to withstand it, even if it does not appear in the meantime, so when it comes again; the structure is already prepared. The slight increase in computing time and the sometimes complex material layout can be cited as drawbacks, which will be dealt with in successive updates. 

Hence, the global damage contribution for each iteration $t$ is the sum of the damage caused by all applied loads $i$, which in turn come from all cyclic instances $j$ for each load, as in Equation \ref{eq:gamma}. 
\begin{equation}
    \\\\\ \gamma^t = \sum_{i=1}^{N_{l,t}}\gamma^t_i = \sum_{i=1}^{N_{l,t}}\sum_{j=1}^{N_{c,i}}\gamma^t_{ij}
\label{eq:gamma}
\end{equation}

\begin{figure}[ht]
    \centering \includegraphics[scale=0.63]{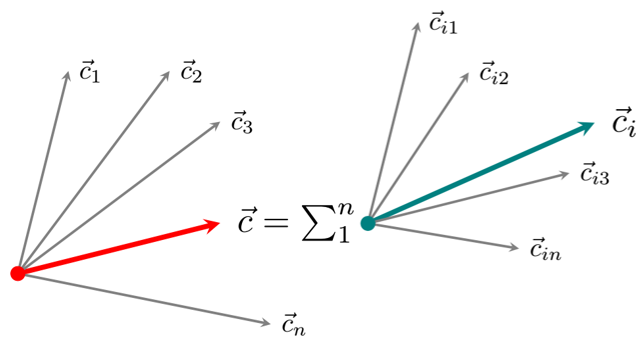}
\caption{\centering Diagram explaining damage contributions in a full iterative loop $t$ as the sum of each load set's contributions $i$.}
\label{fig:Damage_compvol}
\end{figure}

$N_{l,t}$ being the number of different load sets applied in that iteration $t$ and $N_{c,i}$ the loading cycles for each of those loads $i$. Each damage contribution penalizes compliance at every level (cycle $j$, load $i$, iteration $t$) - taking into account the load's probability $p_i$ - as seen in Figure \ref{fig:Damage_compvol} and Equation \ref{eq:compliance}.
\begin{equation}
    \\\\\ c^t = \frac{1}{\gamma^t}\sum_{i=1}^{N_{l,t}}c^t_i = \frac{1}{\gamma^t}\sum_{i=1}^{N_{l,t}}\sum_{j=1}^{N_{c,i}}(p_i c^t_{ij} \gamma^t_{ij}) = \frac{1}{\gamma^t}\sum_{i=1}^{N_{l,t}}\gamma^t_i\sum_{j=1}^{N_{c,i}}(p_i c^t_{ij})
\label{eq:compliance}
\end{equation}

Volume fraction is updated analogously for each load $i$, not at cycle level $j$ like compliance, since the changes in volume are not associated to a cycle probability $p_{ij}$, rather to the load set's influence: 
\begin{equation}
    \\\\\ v^t = \frac{1}{\gamma^t}\sum_{i=1}^{N_{l,t}}v^t_i
\label{eq:volume}
\end{equation}

This volume update accounts for physical wear while avoiding numerical instabilities after embedded density filtering (e.g. null contributions from void elements). The procedure for each iteration is shown in Figure \ref{fig:Damage_compvol} and Algorithm \ref{alg:probabilistic}, using Andreassen's "top88.m" \cite{Andreassen2010} as the main solver).

\begin{algorithm}
\caption{Probabilistic loading}
\begin{algorithmic}[1]
\Require Setup of the model: boundary conditions and bounding volume.
\Ensure \textit{FEA isotropic linear elastic calculation at} \( t = 0 \rightarrow K(^{0}E)U = P \) (top88.m \cite{Andreassen2010})

\While{ \( change  > tol \)}
    \State Initialize \( (\mathbf{c}, \mathbf{dc}, \mathbf{dv}, \mathbf{\gamma}) = \mathbf{0}\)
    
    \For{\( load(i) \in N_{l,t} \)}
        \State \textit{Calculate} \( K, sK \) (top88.m \cite{Andreassen2010})
        \State \textbf{Initialize} \( (\mathbf{c_i}, \mathbf{dc_i}) = \mathbf{0}, \sum p = 0\)
        \For{\( cycle(j) \in N_{c,i} \)}
            \State \textbf{Calculate} \( \mathbf{\gamma_j} \) (see Algorithm \ref{alg:Damage})
            \State \textbf{Update} \textit{$\sum p$} with \textit{p}
            \State \textbf{Update} \( \mathbf{\gamma_i} \) from \( \mathbf{\gamma_j} \) as in Eq. \ref{eq:gamma}
            \State \textbf{Update} \( \mathbf{c_i}, \mathbf{dc_i} \) with \( p, \mathbf{\gamma_j} \) as in Eq. \ref{eq:compliance}
        \EndFor
        \State \textbf{Update} \( \mathbf{c_i}, \mathbf{dc_i} \) with \textit{$\sum p$} as in Eq. \ref{eq:compliance}
        \State \textbf{Initialize} \( \mathbf{dv_i} = \mathbf{1} \)
        \State \textbf{Filter} \( \mathbf{dc_i} \) (sensitivity) and \( \mathbf{dv_i} \) (density)
        \State \textbf{Update} \( \gamma, \mathbf{c}, \mathbf{dc}, \mathbf{dv}  \) with \( \gamma_i, \mathbf{c_i}, \mathbf{dc_i}, \mathbf{dv_i} \) as in Eqs. \ref{eq:gamma} - \ref{eq:volume}
    \EndFor
    \State \textbf{Penalize} \( \mathbf{c}, \mathbf{dc}, \mathbf{dv} \) with \( \gamma \) as in Eq. \ref{eq:gamma}
\EndWhile

\end{algorithmic}
\label{alg:probabilistic}
\end{algorithm}

The nature of cycles $j$ and load probabilities $p_i$ are both determined by the load set's frequency. Load probabilities $p_i$ are computed either directly (if constant or occasional) or by addition of cycle probabilities $p_{ij}$ (when statistic): $p_i = \sum p_{ij}$. Therefore, this framework is probabilistic, as opposed to the deterministic multi-load approach proposed in Section 5 of \cite{Andreassen2010}. For the same iteration $t$, all individual load probabilities add up to 1: $\sum p_i = 1$. 

In practical terms, this cyclic subdivisions only affect statistical loads under assumed distributions. For all three other loading types (constant, alternating and occasional), all cyclic probabilities are considered equal: $p_{ij} = p_i / N_{c,i}$.

In optimization terms, the gradient is now composed not only of the resulting sum of the searching directions for each load $i$, but also each of those loads is the combination of all the weighted gradients for each cycle $j$ - see Equation \ref{eq:compliance} and Figure \ref{fig:Damage_compvol}. This approach accounts for the influence of statistically-measurable uncertainty, multi-axial and multi-load scenarios and fatigue damage considerations, among other criteria. In doing so, it reinforces the evolving topology on the go, as a part of the optimization process itself rather than computationally costly post-processing steps. Some practical examples of Figure \ref{fig:Damage_compvol}'s application can be found in Figure \ref{fig:Cantilever}. 

\begin{algorithm}
\caption{Damage penalization}
\begin{algorithmic}[1]
\For{\( cell \in N_{el} \)}
    \If{\( \rho > 0.8 \)}
        \If {$Fatigue$ criterion} 
            \State \textbf{Calculate} $\mathbf{\gamma_j}$ with $\kappa, \alpha, \sigma$ as in Eq. \ref{eq:Golos}
        \EndIf
        \If {$Traction/compression$ criterion} 
            \State \textbf{Calculate} $\mathbf{\varepsilon}$ from $\mathbf{u}$
            \State \textbf{Find} $\varepsilon_1$
            \If {\( sign(\mathbf{\varepsilon}(cell)) = sign(\varepsilon_1) \)}
                \State \textbf{Calculate} $\mathbf{\gamma_j}(cell)$ with $\gamma_j$ from Eq. \ref{eq:Golos} and $P$ as in Eq. \ref{eq:Crit2_strain}
            \Else
                \State \textbf{Set} $\mathbf{\gamma_j}(cell) = 0$
            \EndIf
        \EndIf
        \If {$Orientation$ criterion} 
            \State \textbf{Calculate} $\mathbf{\varepsilon}$ from $\mathbf{u}$
            \State \textbf{Find} $\vec{r_1}$
            \State \textbf{Calculate} $\mathbf{\gamma_j}(cell)$ with $\gamma_j$ from Eq. \ref{eq:Golos} and $\beta, \vec{R}_{print}$ as in Eq. \ref{eq:crit3_printing} 
        \EndIf
        \If {$Mises$ criterion} 
            \State \textbf{Calculate} $\mathbf{\varepsilon}$ from $\mathbf{u}$
            \State \textbf{Calculate} $\mathbf{\sigma}$ from $\mathbf{\varepsilon}$ as in Eq. \ref{eq:Hooke_generalized}
            \State \textbf{Calculate} $\mathbf{\gamma_j}(cell)$ with $\gamma_j$ from Eq. \ref{eq:Golos} and $\mathbf{\sigma}$ as in Eq. \ref{eq:vonMises_eqstress} 
        \EndIf
    \EndIf
\EndFor

\end{algorithmic}
\label{alg:Damage}
\end{algorithm}

\subsection{Damage Criteria for Material Constraints}

Many damage-based criteria can affect compliance (or whatever target variable to be minimized, e.g. stress \cite{Yang1996}) within a constrained TO framework. The four proposed criteria are damage-based in the sense that the cumulative work $\gamma$ at each iteration penalizes compliance directly, undermining SIMP's assumed isotropic properties in favor of more realistic performances: $c^{t+1}=c^t/\gamma^t$, as in Equation \ref{eq:compliance}. Depending on the TO problem's boundary conditions, the enforcement of these schemes can become rater cumbersome, computationally speaking. 

For that reason, they will only be applied to solid cells ($\rho = 1$) or likely to become so in upcoming iterations ($\rho > 0.8$), to avoid useless evaluation of void ($\rho = 0$) and vanishing elements ($\rho << 1$). See Algorithm \ref{alg:Damage} below expanding Algorithm \ref{alg:probabilistic}'s line $7$ to compute damage criteria. The details of each criterion follow.

\subsubsection{Isotropic Fatigue Damage}

The first suggested criterion features an isotropic energy function \cite{Golos1988,Golos1989} accounting for fatigue damage, containing cyclic ($N_f$) and stress terms ($\sigma _f$). Total damage $\gamma_{ij} = \Delta W_t$ penalizes compliance each iteration $t$ as in Equation \ref{eq:compliance}. For each cycle $j$ of a given load $i$ in iteration $t$
\begin{equation}
    \\\\\ \gamma_{ij} = \Delta W_t = \left[\frac{\kappa}{N_f}\right]^{\alpha} + \frac{1}{2E} \left[\sigma _f\frac{t}{N_f}\right]^2
\label{eq:Golos}
\end{equation}

where $\kappa$ and $\alpha$ are tunable scalars, $N_{f,i}$ the number of loading cycles $j$ for each load $i$, $E$ the base material's Young modulus and $\sigma _f$ the tension damage threshold (another explicit material property). Thus, Equation \ref{eq:Golos} encapsulates both material and loading effects. 

This criterion is isotropic, i.e. it deems the material affected by cumulative wear and tear homogeneously in all its parts and along all directions. This highlights the effects of damage alone in the optimized topological distribution, but other options are also compatible (e.g. crack initiation around notches and stress concentration points). 

This penalization yields results close to the original vanilla configuration of $top88.m$ \cite{Andreassen2010}, while adding reinforcement where needed (e.g. feeble struts and joints prone to stress concentration and thus plasticity and crack initiation). 

\subsubsection{Tunable Mechanic Behavior}

The second criterion concerns the topology's stress state, penalizing each cell's density with the highest principal strain $\epsilon _1$ multiplied by a scalar factor $P$. For all cells undergoing the target strain state (traction/compression):
\begin{equation}
 \\\\\   \gamma = P\Delta W_t  
\label{eq:Crit2_strain}
\end{equation}

Therefore, the resulting structure will be reinforced against the mechanical behaviors the TO process is penalizing. Of course, such responses are strongly dependent on the material's properties, i.e. this criterion is tantamount to material choice, performance-wise: the stress/strain response that suits best the selected material(s) will constitute the designer's preferred choice. 

For instance, compressive states are convenient for materials like plain concrete (with strong compressive properties but negligible traction resistance on its own), whereas traction should be promoted for steel-like materials (preventing buckling), cables (unable to withstand compression), or composites along their fibers.

\subsubsection{Anisotropy}

The third proposed criterion considers more practical constraints. If the material is printed via additive manufacturing (AM), the effects of the prototype's printing direction must be considered, since they have a noticeable impact in their global and local properties, dismantling again SIMP's isotropic assumptions. Anisotropy is present along the printing layers (analogous to fibers in composites) and around defects (scale, porosity, burrs, etc.) \cite{Mueller2018,Zhao2019}. The isotropic damage $\Delta W_t$ is multiplied by the projection $proj$ of the chosen printing direction $\vec{R}_{print}$ onto the first principal direction $\vec{R}_1$ for each cell. 

For unitary vectors:
\begin{equation}
\\\\\ \gamma = \Delta W_t proj(\vec{R}_{print},\vec{R}_1) = \Delta W_t (1+\beta cos(\widehat{\vec{R}_{print},\vec{R}_1}))
\label{eq:crit3_printing}
\end{equation}

where $\beta$ is a custom intensifying scalar parameter. This criterion takes into account the effects of the prototype's manufacturing technique in its mechanical performance. If AM is used, the layer direction will, of course, be the most resistant one, which implies non-homogeneity and anisotropy, affecting the even the failure mode (layer separation rather than plasticity is common behavior in additively printed specimens).



\subsubsection{Plasticity Limit}

Lastly, a fourth option for compliance penalty could be directly von Mises yield criterion, delimiting SIMP's linear elastic assumptions, shear distribution and deformation energy. For that purpose, the stress tensor $\mathbf{\sigma}$ is computed from the Lamé-Hooke equations:
\begin{equation}
    \\\\\ \mathbf{\sigma} = \lambda tr(\mathbf{\varepsilon}) + 2\mu \mathbf{\varepsilon}
    \label{eq:Hooke_generalized}
\end{equation}

$\mu$ being the material's Poisson coefficient. With the stress tensor $\mathbf{\sigma}$, the equivalent von Mises stress $\sigma_{vM}$ can now be computed and used as a damage penalization $\gamma$:
\begin{equation}
    \\\\\ \gamma =\sigma_{vM} = \sqrt{\sigma_{xx}^2+\sigma_{yy}^2-\sigma_{xx}\sigma_{yy}+3\tau_{xy}^2} = \sqrt{\sigma_{1}^2+\sigma_{2}^2-\sigma_{1}\sigma_{2}}
    \label{eq:vonMises_eqstress}
\end{equation}


This criterion provides a quantitative measure for the degree in which linear elasticity is an acceptable approximation or not, as well as giving valuable information on the material's internal distortion. The latter proves especially relevant when anisotropic properties arise, unexpectedly or not. Such is the case of functionally-graded materials (composites, metamaterials). 

Stress is homogenized and so, stress concentrations are softened and plasticity is less likely to occur, effectively shielding the optimized topology from complex non-linear effects. This can be interpreted as a rough compliance-stress TO compromise \cite{Yang1996}.


\section{Results}

In this section, some numerical studies will be showcased to demonstrate this proposal's capabilities and versatility, differentiating between deterministic and probabilistic topologies and applying different penalization criteria at a time. 

\subsection{Probabilistic Topological Optimization}

For nondeterministic conditions, several load types with associated probabilities will be discerned while applying isotropic damage penalization (first criterion, Equation \ref{eq:Golos}) for benchmarking purposes. All examples are shown after 100 iterations for a volume fraction of $v_f = 0.4$, unless specified.

\subsubsection{Conceptual Significance}

\begin{figure}[ht]
\centering \includegraphics[scale=0.50]{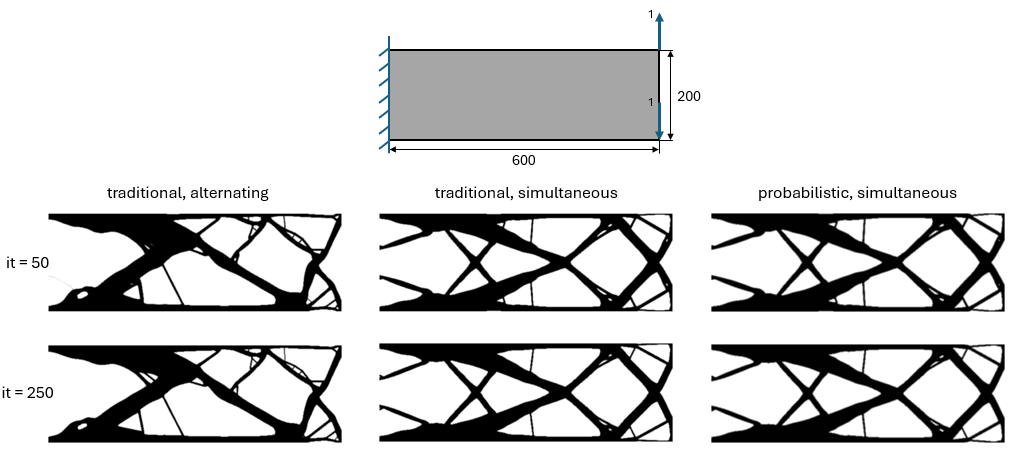}
\caption{\centering Cantilever beam (boundary conditions in top row) and its optimized topologies for the deterministic alternating (left column), deterministic simultaneous (middle column) and probabilistic simultaneous (right column) cases. Iterations 50 (middle row) and 250 (bottom row).
\label{fig:comb_vs_sep}
}
\end{figure}

Firstly, the motivation for the proposed statistical approach must be understood. Let a cantilever beam be considered, with its left side fixed and two unitary outward-facing forces on each right corner, up and down (see Figure \ref{fig:comb_vs_sep} top row). TO is then applied via the $top88.m$ code \cite{Andreassen2010}. 

If both forces are applied simultaneously (constant, static), their compliance contribution is unique under the traditional deterministic approach, thus easing convergence and obtaining a sharp result in few iterations (see Figure \ref{fig:comb_vs_sep} middle column). However, should they be applied alternatively (one each iteration), the result is much more topologically convoluted (Figure \ref{fig:comb_vs_sep} left column): multiple fragile twigs, great asymmetry determined by the alternating load order (first introduced force is dominating), etc. 

All these issues can be tracked down to conflicting optimization directions with different, uncoordinated compliance contributions. More iterations do not help in the second case: iteration 250 displays the same issues as iteration 50, even worse. Importantly, the amount of iterations does not visibly improve the deterministic (vanilla) topologies: no configuration shown along Figure \ref{fig:comb_vs_sep}'s columns reaches noticeably simpler or less compliant solutions.

Should the proposed probabilistic framework (Equations \ref{eq:gamma} and \ref{eq:compliance}) be introduced, with equal probability for each load ($p_{ij}^{up} = p_{ij}^{down} = 0.5$), the result is virtually equivalent to the deterministic case (Figure \ref{fig:comb_vs_sep}  right column), while retaining the same contributions as in the alternate case. This is expected, since the load share is equal for the top and bottom right corners. This is proof of the equivalence of the probabilistic (equal frequency share) and deterministic approaches (equal in modulus) when the prototype is symmetrically loaded.

\subsubsection{Asymmetric Loading}

The solution is straight-forward when symmetric loading is in place, as in the previous example. However, the expectation is not so clear otherwise: if loading is skewed, the result will no longer by symmetric nor easily predictable. The proposed approach (Equations \ref{eq:gamma} and \ref{eq:compliance}) can be used to optimize multi-load cases according to each load's share in modulus (deterministic) or frequency (probabilistic). If isotropic damage penalization (Equation \ref{eq:Golos}) is also introduced, the resulting topology will be more robust and wear-resistant. 

\begin{figure}[ht]
\centering \includegraphics[scale=0.50]{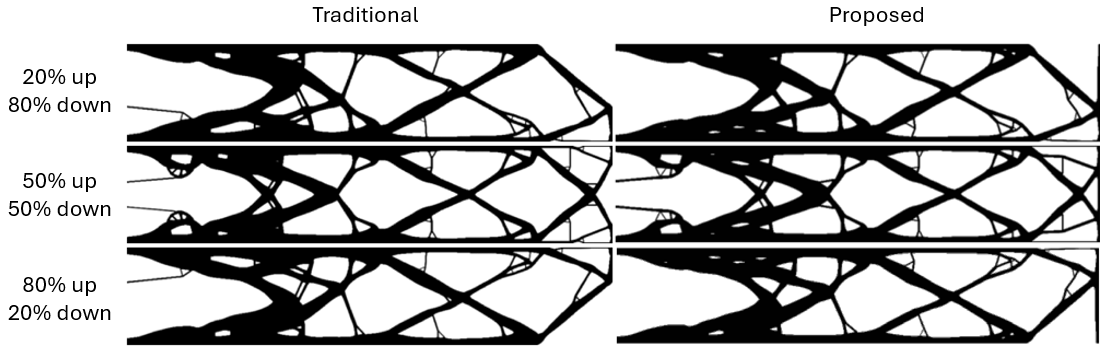}
\caption{\centering Cantilever beam (configuration 1 size 1000x200) loaded as in Figure \ref{fig:comb_vs_sep} top row with different modulus (deterministic, left column) or frequency shares (probabilistic, right column): 20\%-80\% (first row), 50\%-50\% (second row) and 80\%-20\% (third row).}
\label{fig:biload}
\end{figure}

Consider the example cantilever beam in Figure \ref{fig:biload} (same loading conditions as in Figure \ref{fig:comb_vs_sep} top row on a 1000 elements length). It is loaded symmetrically and asymmetrically in frequency, with damage configuration 1 ($\kappa _f=\sigma _f=10^{-5}$ and $\alpha =9\cdot10^{-2}$ in Equation \ref{eq:Golos}). In probablistic examples, the percentages express the fraction of total iteration loops in which each load is applied (frequency share, e.g. 50\%/50\% implies alternate loading, again with $p_{ij}^{up} = p_{ij}^{down} = 0.5$).

\begin{figure}[ht]
\centering \includegraphics[scale=0.40]{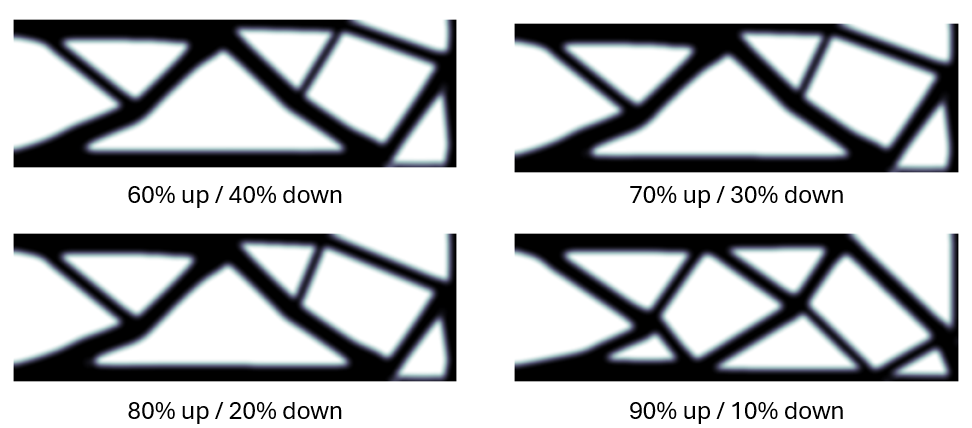}
\caption{\centering Asymmetric loading in frequency, applied to the cantilever beam in Figure \ref{fig:comb_vs_sep} top row with different percentages (up / down): 40\% / 60\% (top left), 30\% / 70\% (top right), 20\% / 80\% (bottom left) and 10\% / 90\% (bottom right).}
\label{fig:cantilever_skewed}
\end{figure}

Although their genus has not been significantly altered, probabilistic designs (Figure \ref{fig:biload} right column) are generally simpler and sturdier than deterministic ones (Figure \ref{fig:biload} left column): absence of thin pins on the left, reinforced struts near supports and loading, etc. Most importantly, only the proposed topologies retain the vertical strut on their right end, sustaining both loads - making them the only practically viable solutions in both skewed cases (20/80, 80/20). 

This proves how the proposed damage-based probabilistic approach not only uses the given target volume fraction more efficiently (joint reinforcement, strategic thickening), but it also keeps structurally relevant parts which would have been lost to the vanilla SIMP process otherwise.

For other frequency combinations in a 600x200 version, see Figure \ref{fig:cantilever_skewed}. When compared to the 50\% / 50\% case (Figure \ref{fig:comb_vs_sep} right column), asymmetrically loaded examples in Figures \ref{fig:biload} and \ref{fig:cantilever_skewed} display, as expected, equally asymmetric topologies with respect to the mid-horizontal axis. As loading share in Figure \ref{fig:cantilever_skewed} is progressively shifted to the bottom half, more volume is diverted to the area to withstand it. 

The least loaded end (top right corner) is still held by the vertical wall along the loading axis, whereas the most loaded point presents an horizontal strut as well, to transfer momentum to the rest of the structure. 

Note how the different beam lengths for the 20\%/80 \% case (1000 in Figure \ref{fig:biload} bottom right and 600 in Figure \ref{fig:cantilever_skewed} top left) yield slightly different features, further accentuated by the diffusion induced by a higher density filtering radius \cite{Andreassen2010} ($r_{min} = 10$ instead of $r_{min} = 10$ like previous examples).

Prior to Figure \ref{fig:cantilever_skewed}, the density filtering radius in \cite{Andreassen2010} was set to $r_{min} = 2$, offering the sharpest possible contrast between material and void. However, these sharp designs come at a cost: they foster very thin twigs which induce fragility (especially in intersections), buckling, artificially increase compliance and pose manufacturing difficulties. 

\begin{figure}[ht]
\centering \includegraphics[scale=0.50]{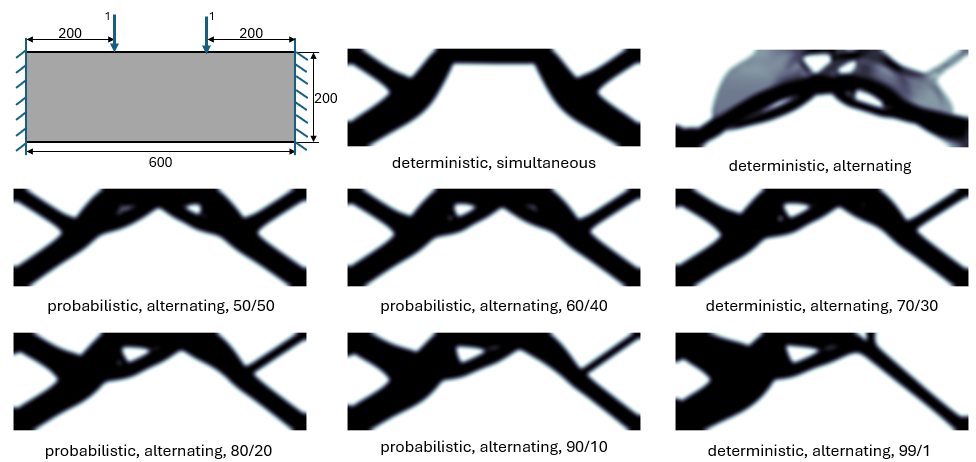}
\caption{\centering 600x200 beams (see top left corner for loading boundary conditions) with various frameworks (deterministic/probabilistic, simultaneous/alternating) and frequency shares (50/50, 60/40, 70/30, 80/20, 90/10, 99/1).}
\label{fig:bifixed_skewed}
\end{figure}

\begin{figure}[ht]
\centering \includegraphics[scale=0.60]{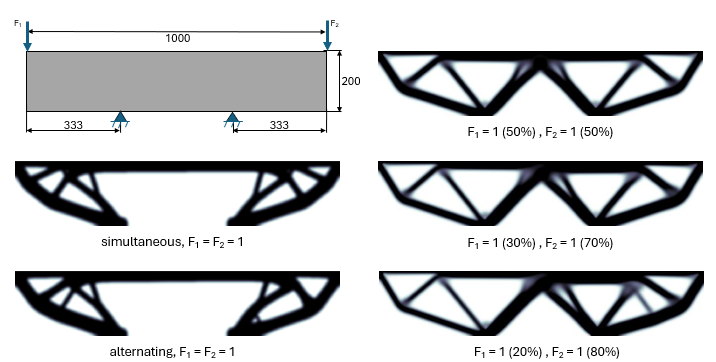}
\caption{\centering 1000x200 beams supported on both ends (boundary conditions pictured top left) in various deterministic (simultaneous/alternating, left column) and probabilistic settings (alternating, right column).}
\label{fig:bisupportf_skewed}
\end{figure}

To alleviate that, the filtering radius can be increased (to $r_{min} = 10$, for instance), obtaining less crisp contours but simpler material layouts with fewer, thicker and structurally significant struts. See Figure \ref{fig:bifixed_skewed} for a supported beam fixed on both walls or Figure \ref{fig:bisupportf_skewed} degrees of freedom on its bottom corners.

Figure \ref{fig:bifixed_skewed} shows how volume is redistributed according to loading frequencies, although not proportionally, since the final topology is also conditioned by the underlying fixed boundary conditions. In the cantilever examples (Figures \ref{fig:comb_vs_sep} and \ref{fig:biload}), the structure was globally isostatic and so the right part was free to develop within the algorithm's prescriptions. 

Conversely, the boundary conditions in Figure \ref{fig:bifixed_skewed} - both walls fixed on both degrees of freedom - make the resulting structure hyper-static, and thus much more restricted. As loading shifts progressively to the left side, so does the volume distribution in the optimized topologies (Figure \ref{fig:bifixed_skewed} middle and bottom rows): material agglomerates on the left half, gradually breaking symmetry to the point of erasing the top right pin in the most unbalanced case (99/1, bottom right corner). 



A similar array  of case studies for a centrally bi-supported beam with loads on both top corners is contained in Figure \ref{fig:bisupportf_skewed}. Since the supported example (Figure \ref{fig:bisupportf_skewed}) is less constrained than its fixed counterpart (Figure \ref{fig:bifixed_skewed}), the deterministic cases (left column) are quite similar. However, simultaneous loading (center left) yields axial symmetry with respect to the middle vertical axis, whereas alternating loads (bottom left) do not, allocating slightly more volume on the right side (loaded first). 

For these particular boundary conditions, the probabilistic alternatives (Figure \ref{fig:bisupportf_skewed} right column) are quite different from the deterministic ones. 

By optimizing the resulting cumulative gradient of each frequency-weighted load instead of fully applied and constant, the probabilistic designs are topologically simpler, distributing volume more efficiently and protecting the central section between supports. 
Up to this point, skewed loading has only involved different frequencies in load application, but skewed moduli have similar effects and can be used as an alternative or in a combined way.

\begin{figure}[ht]
\centering \includegraphics[scale=0.40]{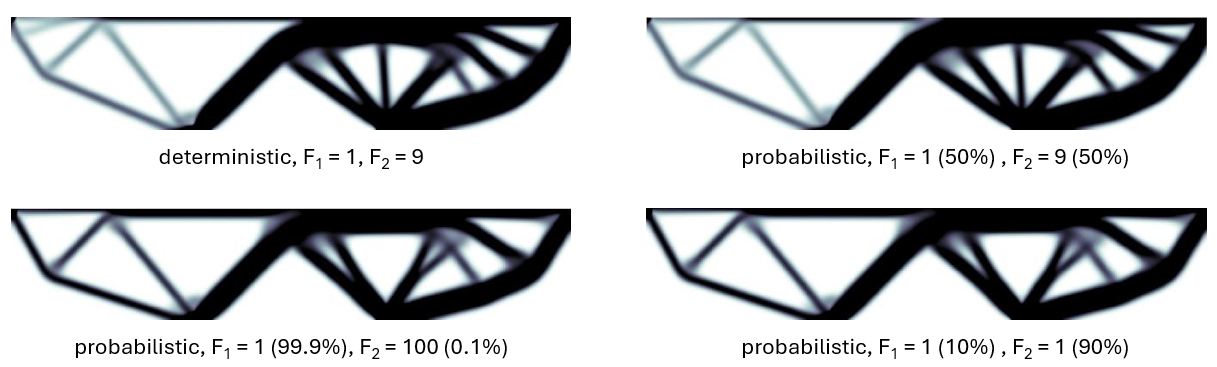}
\caption{\centering 1000x200 beams with different load frequencies and moduli. Boundary conditions in Figure \ref{fig:bisupportf_skewed} top left.}
\label{fig:bisupportf_skewed_moduli}
\end{figure}

Consider Figure \ref{fig:bisupportf_skewed_moduli} displaying practically identical topologies row-wise. The top row proves yet again that deterministic and probabilistic approaches are equivalent when loading frequencies are symmetric - as in Figure \ref{fig:comb_vs_sep}, while the bottom row demonstrates how the proposed probabilistic loading framework can be interchangeably enforced via frequency or modulus variation, as long as their compliance contributions are identical.

Applying constant and simultaneous loading in each iteration (deterministic, Figure \ref{fig:bisupportf_skewed_moduli} top left) is tantamount to doing so alternatively with a 50\% chance each. If this modulus imbalance is partially compensated by a countering frequency bias (Figure \ref{fig:bisupportf_skewed_moduli} top left), the resulting layout is equivalent to that obtained by assigning equal loading in modulus and adjusting the frequencies accordingly, as in Figure \ref{fig:bisupportf_skewed_moduli} top right.

\subsubsection{Multi-axial Loading}

\begin{figure}[ht]
\centering \includegraphics[scale=0.50]{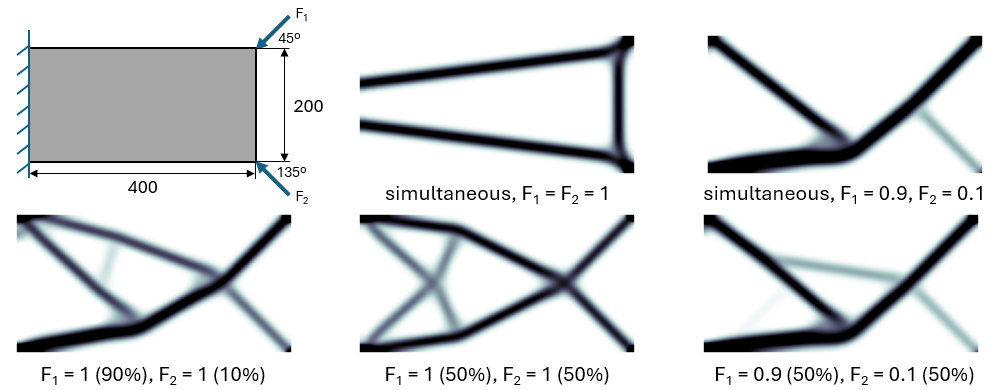}
\caption{\centering Tilted loads (loading conditions: top left) on a 400x200 cantilever beam with $v_f = 0.2$. Iteration 100. Deterministic simultaneous (top row) and vs. probabilistic cases (bottom row).
\label{fig:tilted}
}
\end{figure}

Multi-axial optimization with different loading types is also feasible under this framework. See Figure \ref{fig:tilted} for comparison purposes, regarding the cantilever beam case.

Very different topologies are obtained for the symmetrically loaded case, i.e. equal share in modulus and frequency (middle column). 

While simpler, the deterministic (top middle) design's load transmission is weaker and vulnerable to small variations, whereas the probabilistic alternative (bottom middle) is more robust as volume is better distributed - i.e. covering wider angle ranges. 

Such is the case for asymmetric loading as well (Figure \ref{fig:tilted} right column): the overall volume layout is similar, but the probabilistic case (bottom right) is again more robust (additional middle strut for better load transmission) than its deterministic twin (top right).

A cantilever 600x200 beam is loaded in multiple directions: top right ($pos_x = 600$) and bottom middle ($pos_x = 300$) with two static leftward horizontal loads (5 N) applied every 20 loading cycles and two alternating vertical loads varying in modulus and position via two distinct normal distributions: $\mathcal{N}(1,0.5)$ for modulus and $\mathcal{N}(pos_x,50)$ for position - 10 cyclic samples are drawn per iteration. 

\begin{figure}[ht]
\centering \includegraphics[scale=0.50]{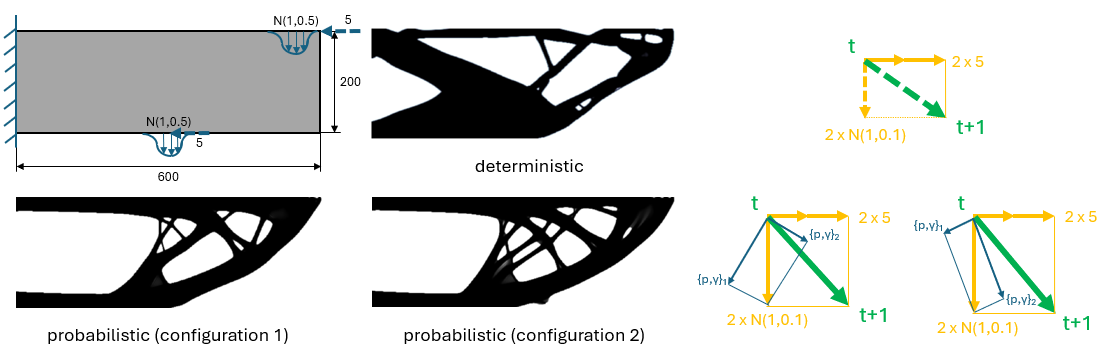}
\caption{\centering Multi-load cantilever topologies (filter radius $r_{min}=2$) with load position and modulus uncertainty (top left): vanilla (top middle), configuration 1 with $\kappa _f=\sigma _f=10^{-5}$ and $\alpha =9\cdot10^{-2}$ (bottom left) and configuration 2 with $\kappa _f=5\cdot10^{-5}$, $\sigma _f=5\cdot10^{-2}$ and $\alpha =9\cdot10^{-3}$ (bottom middle), along their respective load diagrams (right) at the cycle (blue), load (yellow) and loop level (green).
\label{fig:Cantilever}
}
\end{figure}

See Figure \ref{fig:Cantilever} top left for the boundary conditions. The regular approach (top middle) is unrealistic and feeble for distributed loading, clearly sub-optimal: unadapted to the uncertainty induced by the distributed and impact loads, prone to buckling and quite oversized, resulting in a poor volume distribution: excessive for the bottom middle and insufficient for the top right end. 

Conversely, the modified versions using the proposed methodology (Figure \ref{fig:Cantilever} bottom row) offer more organically-shaped, intuitive designs where the topology actually reinforces struts linking loaded points between each other (e.g. bottom right curve) in a direct and smooth manner (thus, less prone to stress concentration). The probabilistic alternatives harness loading by strengthening (left pins and overall right contour), adding (middle region) and/or bifurcating struts (right end) in several directions (wider direction range for loading). 

Loading diagrams in Figure \ref{fig:Cantilever}'s right side shed some light on the obtained topologies. In the deterministic case (top row), the contribution of constant loads is well accounted for (solid yellow arrows), whereas the distributions' (dashed yellow arrow) changes with every iteration (thick, dashed green arrow). The latter provokes an ill-defined gradient that thwarts convergence and so yields impractical topologies (Figure \ref{fig:Cantilever} top middle). 

In the probabilistic examples, however, every iteration is well defined from the cyclic level $j$ (thin, blue arrows) - with associated probabilities $p_{ij}$ and damages $\gamma_{ij}$, through the load level $i$ (yellow) and finally adding up to iteration $t$ (green). Hence, although cyclic contributions remain randomized and unpredictable, each iteration has an univocal, well-defined gradient for each iteration loop, i.e. weighted gradient descent. Notably, using a standardized statistical distribution with known mean and deviation allows for reliable confidence intervals, widespread in industry.

Alas, two main drawbacks are also patent in Figure \ref{fig:Cantilever}: topologies are more convoluted (higher genus) - which could entail manufacturing issues - and several intermediate-density pockets arise inbetween struts. This can be erased via filtering \cite{Sigmund2007} - with the consequent increase in computational time. Both inconveniences can be partially dealt with via clever choice of parameters in Equation \ref{eq:Golos}. 

\begin{figure}[ht]
\centering \includegraphics[scale=0.50]{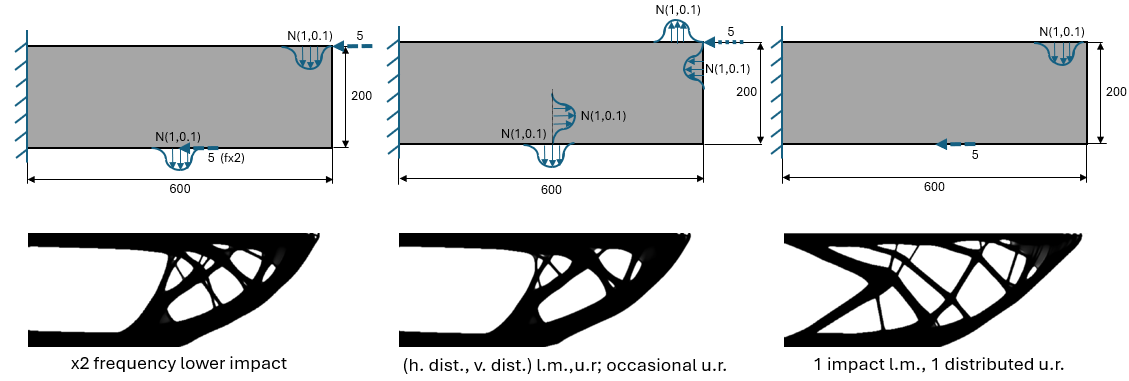}
\caption{\centering Variations on Figure \ref{fig:Cantilever} (with configuration 1): bottom impact gets doubled in frequency (left), loads distributed vertically and horizontally plus an occasional horizontal load on the top right corner (middle) and only one vertically distributed load and one horizontal impact (right).
\label{fig:variations}
}
\end{figure}

Interestingly, if some of the loads in Figure \ref{fig:Cantilever} are modified (in modulus, frequency and/or position) or outright erased, the resulting probabilistic topologies remain somewhat similar (e.g. same overall contour), hinting at the existence of some “canonical” solution, perhaps globally optimal. Self-evidently, this is more visible for small loading variations. Consider Figure \ref{fig:variations} for some examples. For the same iteration (100), Figure \ref{fig:variations}'s similarities with the solutions in Figure \ref{fig:Cantilever} are clearer when the loading conditions undergo slight changes. Different material layouts fill up the space inside the outer “shell” (contour) depending on loading conditions. Of course, what constitutes a convenient solution is up to the designer's requirements (material, volume fraction, type of loading, etc.). 

For instance, making horizontal impact loads follow the same distribution as the vertical ones - and adding an occasional impact load of 10 N at iteration 50 - (Figure \ref{fig:variations} middle) barely alters the design seen in Figure \ref{fig:Cantilever} middle. In contrast, just doubling the frequency of one of the applied loads (horizontal impact in the bottom middle now happens each 10 iterations, Figure \ref{fig:variations} left) creates a sort of “hybrid” topology with some features of Figure \ref{fig:Cantilever} bottom left and others from Figure \ref{fig:Cantilever} bottom right. 

Despite all these variations in boundary conditions, the overall contour remains virtually unchanged. 
Remarkably, this design is similar to that of cancellous bone tissue, as they both seek to be optimized for the most varied eventual loads in all directions and moduli, i.e. quasi-isotropic. In a Machine Learning scenario, this would be tantamount to a validation step: the topology which has been optimized for a known set of testing loads can now withstand previously unseen loading with minor structural changes.

\begin{figure}[ht]
\centering \includegraphics[scale=0.50]{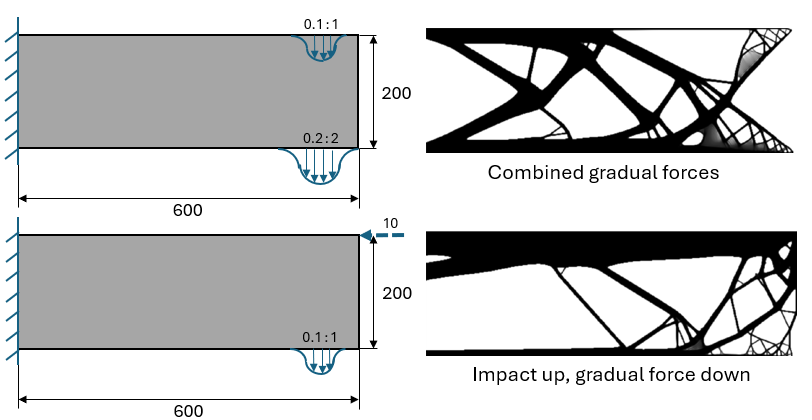}
\caption{\centering Variations on results from Figures \ref{fig:Cantilever} and \ref{fig:variations} (with Configuration 1), applying gradual forces on both corners (top row) or one in the bottom and an impact on top (bottom row). Iteration 500.
\label{fig:gradual}
}
\end{figure}

Since damage penalization (Equation \ref{eq:gamma}) comes with each iteration, applying forces gradually should affect the topology progressively as well, which could be more representative of some scenarios in the service life of a prototype, e.g. loading and unloading. Two cantilever examples are examined in Figure \ref{fig:gradual}. 

In both cases, loading starts with a 10\% of the original value, increasing by a 10\% in each iteration until it reaches 100 \%. Then, it follows the same process in descending order, back to the starting point. This loading-unloading cycle is continuously repeated until convergence.

Figure \ref{fig:gradual} top row combines gradual distributed vertical forces with different modulus and positions: $|F| = \mathcal{N}(1,0.1), |x| = \mathcal{N}(575,25)$ on the top right corner and $|F| = \mathcal{N}(2,0.2), |x| = \mathcal{N}(550,50)$ on the bottom one. They are always present (simultaneous), while having different iterative moduli and positions. The generated branches are thicker and broader in the bottom right corner than in the top one, since the top load is twice as big in modulus and twice as spread in position. Its deeply ramified ending again resemble trabecular bone (as Figure \ref{fig:Cantilever} right).

Figure \ref{fig:gradual} bottom row features an occasional horizontal impact load ($|F| = 10, |y| = 200$) - applied every 10 iterations - on this gradual loading ($|F| = \mathcal{N}(1,0.1), |x| = \mathcal{N}(575,25)$ on the bottom right corner). Heavy ramification is present too, but truncated due to the bigger impact load. When the impact comes for the first time, it displaces a fair share of material to withstand it, neglecting the bottom distributed load whose scattered needs are only partially met. This imbalance shows the importance of eventual non-service loads (e.g. impacts), ignored by deterministic TO. 

This gradual introduction of (un)loading leads to strut ramifications: as the algorithm progresses, it distributes material where it is first needed. As the force is increased in modulus and varied in position, such branches become thicker in order to withstand the ever-growing load while accounting for its spatial dispersion. The inverse is true for unloading: tweaks become thinner, but not excessively, since they still have to cover a wide range of loading points. This way, the algorithm implicitly ensures a minimum threshold thickness for all generated struts. 

\subsubsection{Manufacturability}

\begin{figure}[ht]
\centering \includegraphics[scale=0.50]{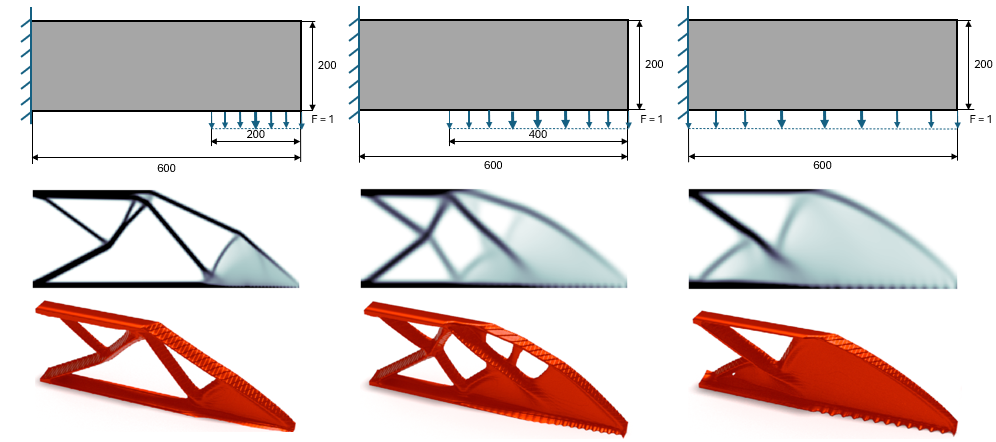}
\caption{\centering Cantilever beams (600x200) loaded with constant unitary modulus and spatial distributions $\mathcal{N}(500,100)$ (left column), $\mathcal{N}(400,200)$ (middle column) and $\mathcal{N}(300,300)$ (right column). Iteration 250. Suggested 3D representations can be found below, reinterpreting density as thickness.}
\label{fig:cantilever_dist}
\end{figure}

For a spatially-undetermined setting, consider a 600x200 cantilever beam fixed on its left side where downward loading is distributed around the bottom right corner. Such loads feature a constant unitary modulus, whereas their position is randomly drawn from a normal distribution $\mathcal{N}(\mu,\sigma)$ for 25 cycles per iteration - $\mu$ being its mean and $\sigma$ its standard deviation. Cases for spatial distributions $\mathcal{N}(500,100)$, $\mathcal{N}(400,200)$ and $\mathcal{N}(300,300)$ are displayed in Figure \ref{fig:cantilever_dist}, showing how position-wise uncertainty in loading yields diffuse material layouts with sizable intermediate density pockets (blurry areas). 

The broader the load distribution, the greater the share of these intermediate pockets. Since these intermediate densities are still stiffer than void, compliance is lower when loading is more sparsely distributed. For instance, the design in Figure \ref{fig:cantilever_dist} center is about 1.5 times stiffer than Figure \ref{fig:cantilever_dist} right, i.e. in the same proportion as their loading lengths. This eases compliance minimization, despite partially defeating the purpose of binary material-void results, which is always subject to filtering anyways.

\begin{figure}[ht]
\centering \includegraphics[scale=0.67]{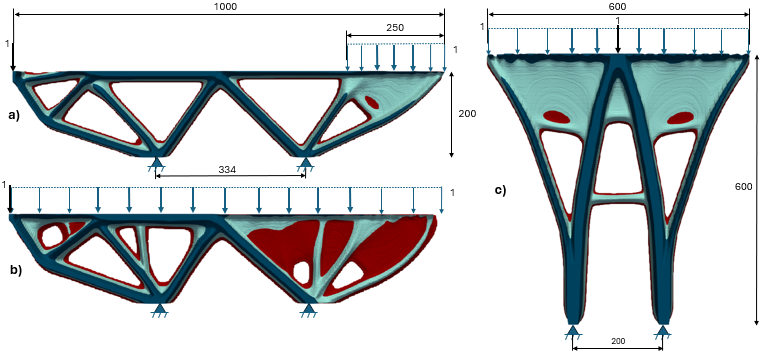}
\caption{\centering Different combination of punctual (black arrows) and randomly-distributed (blue arrows) constant loads on beams with $v_f = 0.2$. Dark blue represents maximum thickness, light blue 50\% of the maximum and red its 10\%.}
\label{fig:bisupport_dist_3D}
\end{figure}

Intermediate density areas can be interpreted as a thickness gradient in 3-dimensional manufactured parts, as in Figure \ref{fig:bisupport_dist_3D}. This results in two direct advantages. Firstly, accepting intermediate densities instead of a sharp material-void layout speeds up the optimization process, since convergence is now reliant on a design criterion (desired compliance) on top of minimization. Secondly, breaking the SIMP method's assumed isotropy by defining density gradients (and thus, stiffness gradients) holds great potentiality for functionally graded behavior, as in composites and metamaterials (e.g. bone implants). 

Figure \ref{fig:bisupport_dist_3D} shows several non-binary topologies where material is distributed by layers, imitating level-set optimization techniques \cite{Wang2003}. The effects of deterministic loading (constant in modulus, position and frequency) and randomly-spaced loads can be seen. 

In all three examples, the “master struts” with maximum stiffness (dark blue) are given by the constant loads (black arrows), while random spatial distributions dictate the intermediate density areas (light blue, red). 

As expected, symmetric loading implies mostly symmetric topologies, although not exactly due to the induced randomness in loading position (Figure \ref{fig:bisupport_dist_3D}c top left). Remarkably, the smaller the randomly-distributed loading is, the more closely it resembles punctual loading, due to momentum balance. 

That explains why Figure \ref{fig:bisupport_dist_3D}a, featuring a reduced distributed loading on one end balancing the sole punctual load in the other, is more symmetric than the fully randomly-loaded case (Figure \ref{fig:bisupport_dist_3D}b). This also applies to Figure \ref{fig:cantilever_dist}.

\subsection{Multi-Objective Topological Optimization}

The proposed damage penalization approach can now be applied to more practical aims on an element basis, such as fatigue, mechanical response, manufacturing concerns and distortion. Again, $top88.m$ \cite{Andreassen2010} will be the base solver for 2D examples. The proposed damage criteria will be enforced under deterministic, static conditions to ponder their individual effects, then combined in synergic ways and ultimately applied to some nondeterministic loading scenarios.

\subsubsection{Individual Criteria}

\begin{figure}[ht]
\centering \includegraphics[scale=0.50]{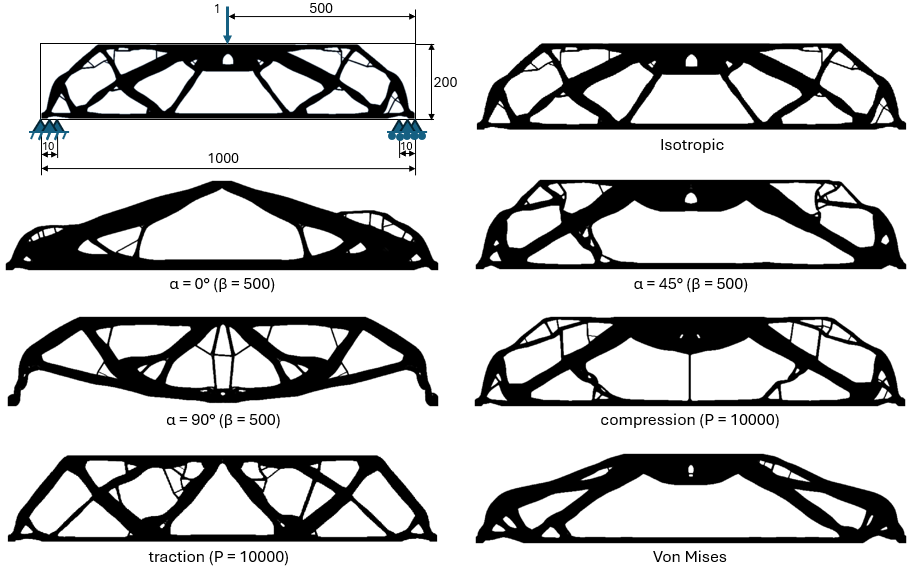}
\caption{\centering 3-point bending beam (size 1000x200). Row-wise, left to right: deterministic and boundary conditions, isotropic (configuration 3 with $\kappa _f=5\cdot10^{-7}$, $\sigma _f=5\cdot10^{-3}$ and $\alpha =-9\cdot10^{-5}$), printing angles of 0$^{\circ}$, 45$^{\circ}$ and 90$^{\circ}$ (all three with $\beta = 500$), compression, traction (both with $P=10000$) and von Mises equivalent stress criteria.
}
\label{fig:3PB_compcriteria}
\end{figure}

Picture a 3-point bending test beam (Figure \ref{fig:3PB_compcriteria} top left) subjected to a central downward unitary load. Strong supports (10-element wide) are placed for more feasible boundary conditions in experimental test settings. As a result, global isostaticity is locally compromised. Upon isolated application of each individual criterion, some optimized topologies can be found in Figure \ref{fig:3PB_compcriteria}. In this section, iteration 200 for $v_f = 0.4$ is shown unless said otherwise.

The isotropic criterion (Figure \ref{fig:3PB_compcriteria} top right) is virtually equivalent to the vanilla $top88.m$ case (top left), except for a few reinforcements on strut junctions and near the supports. It can be considered a reference for comparison.  

\newpage

As expected, the printing criterion reinforces struts along the chosen direction, whether horizontally (0$^{\circ}$, Figure \ref{fig:3PB_compcriteria} second row left), obliquely (45$^{\circ}$, Figure \ref{fig:3PB_compcriteria} second row right) or vertically (90$^{\circ}$, Figure \ref{fig:3PB_compcriteria} third row left), following the layer's direction or the closest angle they can reach for a feasible solution while withstanding the applied loads. 

Note how, for fully horizontal (0$^{\circ}$) or fully vertical (90$^{\circ}$) printing directions, some symmetry with respect to the vertical loading axis is expected. For sideways angles (30$^{\circ}$, 45$^{\circ}$, 60$^{\circ}$, etc.), this symmetry has to be explicitly enforced by mirrored optimization with respect to the loading axis.

\begin{figure}[ht]
    \centering
    \includegraphics[width=0.67\linewidth]{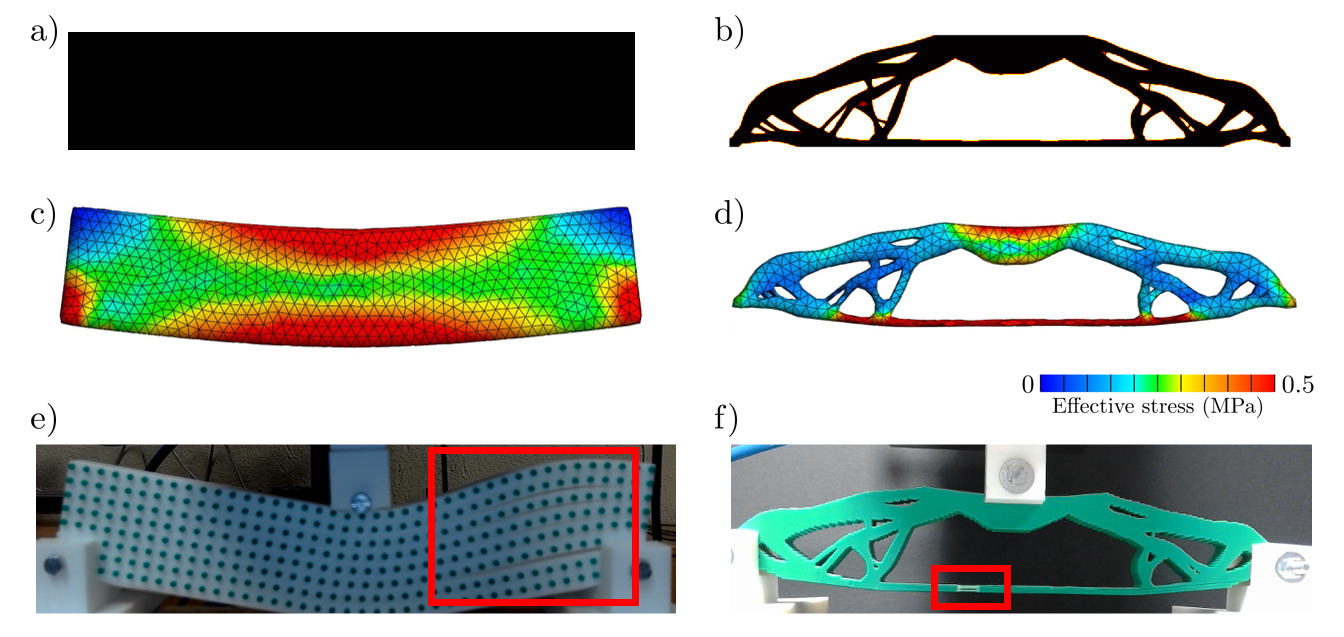}
    \caption{\centering 3-point bending beam (Figure \ref{fig:3PB_compcriteria} top left) as a bulk (left column) and compliance-minimized (SIMP) with von Mises stress penalization (right column): material layout (first row), simulated response via Finite Elements (middle row) and experimental testing (bottom row). Failure mode zoomed in red squares.}
    \label{fig:3PB_vM_vs_0deg}
\end{figure}

Compression (Figure \ref{fig:3PB_compcriteria} third row right) and traction (Figure \ref{fig:3PB_compcriteria} bottom left) penalties act analogously for mechanical responses. They favor compression or traction responses, respectively, by distributing material along the directions in which those responses are expected, with the necessary topological complexity to maximize the number of “struts” undergoing the desired response type. 

The distortion criterion (von Mises, Figure \ref{fig:3PB_compcriteria} bottom right) reduces elastic deformation (and thus, compliance) by rearranging the material mostly along the two diagonals connecting the load application point in the top middle to the supports on the bottom left and right corners, hence minimizing distortion energy. 

Interestingly, von Mises penalization can eliminate layer orientation dependency, thus changing the topology's failure mode. On top of the most inefficient volume distribution in Figure \ref{fig:3PB_vM_vs_0deg} ($v_f=1$ in $a$ vs $v_f=0.4$ in $b$), stress is globally higher in the bulk design ($c$) than in the optimized version ($d$). Despite both being additively printed in horizontal layers, the failure mode differs: layer separation - printing - ($e$) and plasticity ($f$) - material -, respectively.

Some patterns can be observed in Figure \ref{fig:3PB_compcriteria}. Isotropic (top right), 90$^{\circ}$ printing (third row left) and traction cases (bottom left) share a master layout, resembling an X-truss. Similarities can also be spotted for 0$^{\circ}$ (second row left) and von Mises (bottom right); and the isotropic (top right) and 45$^{\circ}$ (second row right) cases: a big triangle joining loading and supports and some inner struts with varying thicknesses. 

A slight topological asymmetry is also observed in Figure \ref{fig:3PB_compcriteria}, due to asymmetric boundary conditions required to ease experimental testing: the supports on the bottom corners are not exactly equal (the left end is more constrained than the right one, to keep global isostatic conditions). The fixed support on the left creates local hyperstaticity, despite the structure being globally isostatic.


Figure \ref{fig:3PB_compcriteria} shows how auxiliary thin tweaks are generated to reinforce thicker and stronger struts. Unfortunately, although mathematically optimal, such tweaks are not structurally meaningful, in turn provoking unnecessary deformation (thus increasing compliance) and being difficult to manufacture and fragile. Two possible solutions exist to alleviate this problem: either decreasing the target volume fraction $v_f$ or applying filters (like the density-based one in \cite{Andreassen2010}), be it during TO or as an extra post-processing step. 

\begin{figure}[ht]
\centering \includegraphics[scale=0.50]{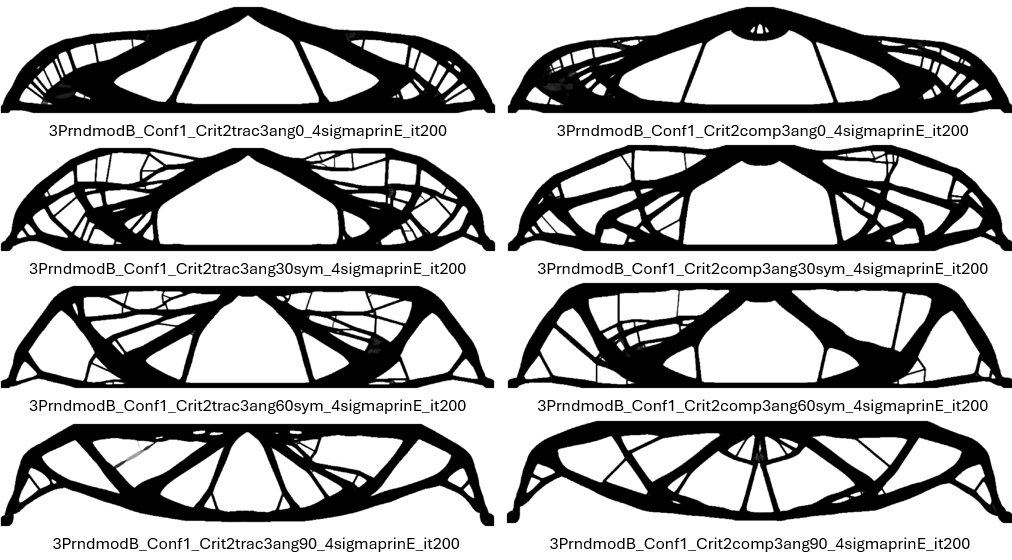}
\caption{\centering Figure \ref{fig:3PB_compcriteria} top left under simultaneous criteria: von Mises stress, traction (left column) or compression (right column) and printing angles 0$^{\circ}$ (first row), 30$^{\circ}$ (second row), 60$^{\circ}$ (third row) and 90$^{\circ}$ (fourth row).}
\label{fig:3PB_allcriteria}
\end{figure}

\begin{figure}[ht]
\centering \includegraphics[scale=0.65]{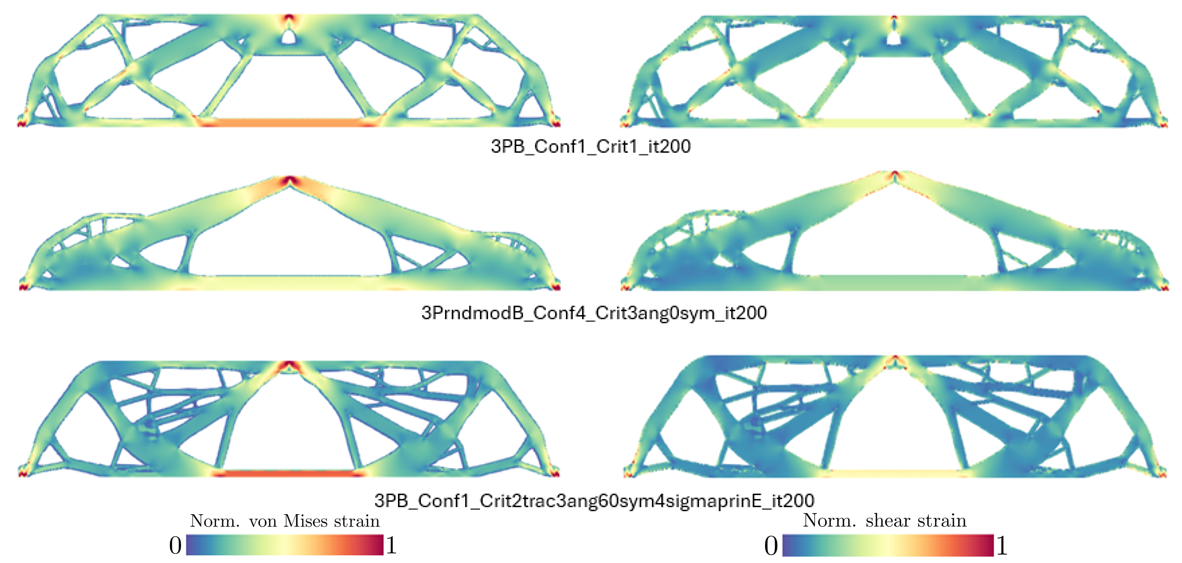}
\caption{\centering Figure \ref{fig:3PB_compcriteria} top left under various combined criteria. Normalized mechanical response (left: von Mises equivalent stress, right: shear strain) under isotropic damage (first row), printing angle 0$^{\circ}$ and von Mises (second row) and traction, printing angle 60$^{\circ}$ and von Mises (third row).}
\label{fig:3PB_vM_shearstrain}
\end{figure}

\begin{figure}[ht]
\centering \includegraphics[scale=0.55]{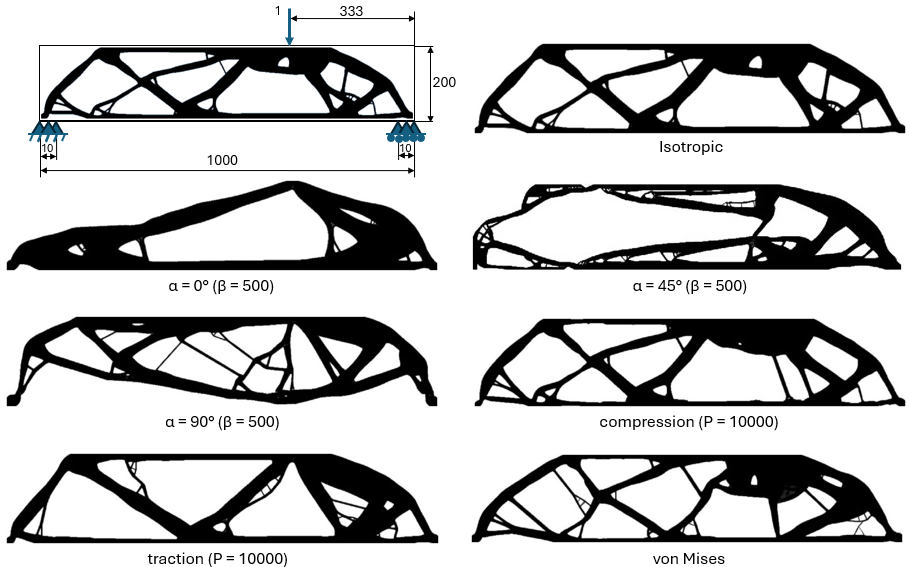}
\caption{\centering Examples from Figure \ref{fig:3PB_compcriteria} with rightwards-skewed loading at two thirds of their length.}
\label{fig:3PB_compcriteria_skewed}
\end{figure}

\subsubsection{Combined Criteria}

While the proposed criteria can in fact be combined (mostly without mutual canceling effects), some of them are prevalent over others, which must be accounted for when facing multiple design requirements. In Figure \ref{fig:3PB_allcriteria}, several criteria were applied simultaneously to a 3-point bending beam with a load whose modulus is given by the same boundary conditions as Figure \ref{fig:3PB_compcriteria} top left. A naming convention of the form $Load.ConfA.CritXYZ.itIII$ has been adopted to classify topologies, denoting: loading conditions, isotropic parameter configuration (Equation \ref{eq:Golos}), applied criteria with their respective specifications and iteration number ($200$, in this case).

All examples in Figure \ref{fig:3PB_allcriteria} obey the von Mises criterion, minimizing distortion by avoiding unnecessary ramifications. However, the topologies only resemble the isolated canon (Figure \ref{fig:3PB_compcriteria} bottom right) when combined with a 0$^{\circ}$ printing direction (first row) - also similar to the isotropic form (Figure \ref{fig:3PB_compcriteria} top row). Designs penalized for traction (Figure \ref{fig:3PB_allcriteria} left column) mimic their canonical shape (Figure \ref{fig:3PB_compcriteria} bottom left) when the printing angle is high, i.e. 60$^{\circ}$ (Figure \ref{fig:3PB_allcriteria} third row left) and 90$^{\circ}$ (Figure \ref{fig:3PB_allcriteria} fourth row left), being also similar to the 90$^{\circ}$ simple criterion (Figure \ref{fig:3PB_compcriteria} third row left). The opposite can be said for the compression (Figure \ref{fig:3PB_allcriteria} right column): the smaller the printing angle is, the more similar the topologies are to the pure compression canon (Figure \ref{fig:3PB_compcriteria} third row right). Therefore, some combinations of criteria are synergic while others dilute some of their effects. The combined effects of various criteria at once promotes thin tweaks inbetween thicker struts. 


\subsubsection{Mechanical Response}

Out of all possible results obtainable through post-processing, shear strain and von Mises equivalent stress are deemed the most relevant to ponder the criteria's consequences on mechanical response, since they yield the most complex gradients with stress concentration in loading points and supports. For a meaningful comparison, results coming from different examples are normalized ($\mathbf{x}_{norm} = (\mathbf{x}-\bar x)/\sigma_x$).

In Figure \ref{fig:3PB_vM_shearstrain}'s top row, the plain isotropic topology (Figure \ref{fig:3PB_compcriteria} top right) exhibits strain and stress concentrations around loading and support points, most strut junctions and along the center bottom strut. 

For the printing criterion with 0$^{\circ}$ (middle row, Figure \ref{fig:3PB_compcriteria} top right), stress and strain concentrations lessen in the bottom strut in exchange for a slightly more stressed area around the loading point. For 60$^{\circ}$ under traction penalty (bottom row, Figure \ref{fig:3PB_allcriteria} third row left column), the situation is close to the isotropic case, with smaller stress and strain concentrations in junctions. 

Which combination is preferable depends greatly on the material's mechanical properties, loading conditions and the designer's goals and choices (e.g. reinforcing highly stressed areas with thicker layers and/or a more resistant material). Note how the middle and bottom topologies have been slightly filtered to reduce their topological complexity.

\subsubsection{Asymmetric and Restrictive Boundary Conditions}

Should loading conditions be asymmetric, the topologies will be too. Consider shifting the load in Figure \ref{fig:3PB_compcriteria} rightwards to two thirds of the beam's total length, as in Figure \ref{fig:3PB_compcriteria_skewed} top left. The same isolated criteria are applied to such scenario. The most loaded half of the structures (right) in Figure \ref{fig:3PB_compcriteria_skewed} is clearly more reinforced (thicker struts, more volume concentration), while the left half is dominated by thinner, longer struts connecting the distant load to the corresponding support. This leaves room for functional grading, (tailored anisotropy), although some mechanical issues could arise (e.g. buckling on the left side).

\begin{figure}[ht]
\centering \includegraphics[scale=0.50]{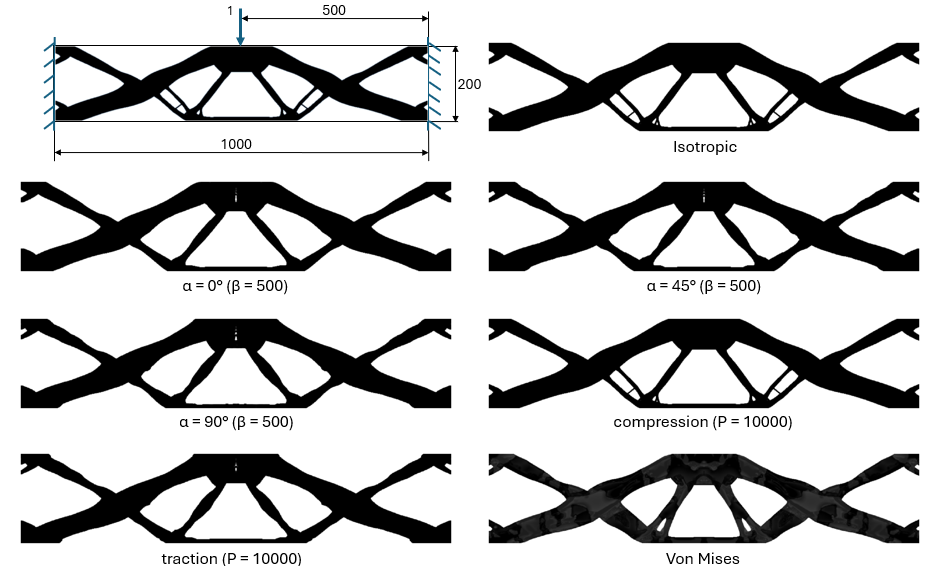}
\caption{\centering Fixed beam examples for single isolated criteria. The highly restricted configuration forces all the cases to converge to roughly the same configuration}
\label{fig:3PBfixed_compcriteria}
\end{figure}

The optimized topologies will not be as diverse if boundary conditions are too restraining (e.g. hyperstaticity). This gives less freedom for topological evolution, resulting in most criteria being either slight variations of a constrained optimum or ill-defined density maps. See Figure \ref{fig:3PBfixed_compcriteria} for an example of an hyperstatic setting. The almost identical topologies for all proposed criteria in Figure \ref{fig:3PBfixed_compcriteria} demonstrates how certain boundary conditions are too restrictive, rendering damage penalization somewhat powerless. Similarity between the isotropic (top right) and compression cases (third row right) still applies. The von Mises stress criterion (bottom right) yields an irregular material distribution, jeopardizing convergence.

\subsubsection{Damage Criteria under Nondeterministic Loading}

\begin{figure}[ht]
\centering \includegraphics[scale=0.40]{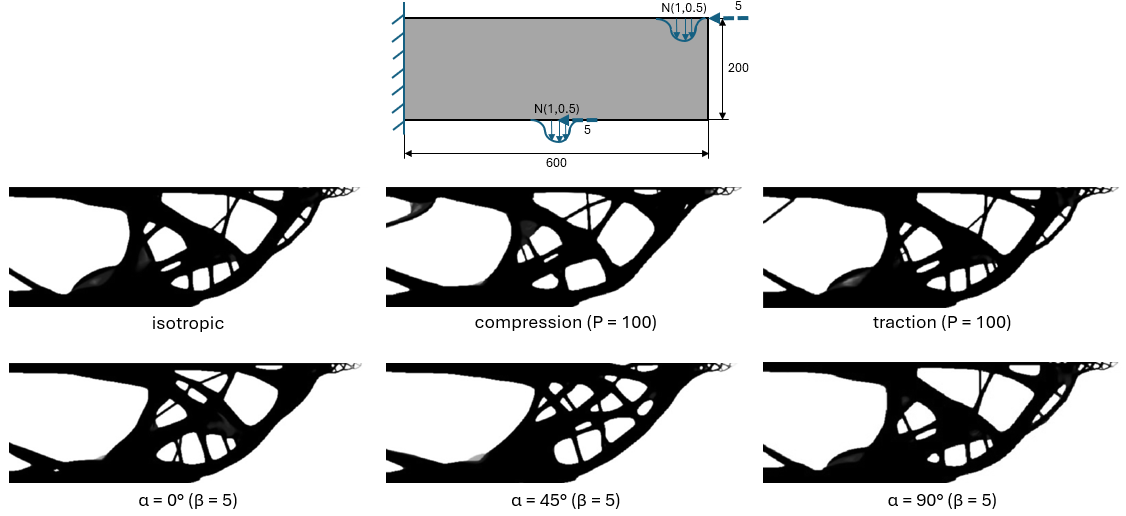}
\caption{\centering Cantilever beam (600x200) with uncertain loading conditions (top row). Row-wise, from left to right: isotropic (configuration 2 with $\kappa _f=5\cdot10^{-5}$, $\sigma _f=5\cdot10^{-2}$ and $\alpha =9\cdot10^{-3}$), compression and traction (both with $P=100$) and printing angles of 0$^{\circ}$, 45$^{\circ}$ and 90$^{\circ}$ (with $\beta = 5$). Iteration 100.
}
\label{fig:cantilever_low}
\end{figure}

\begin{figure}[ht]
\centering \includegraphics[scale=0.40]{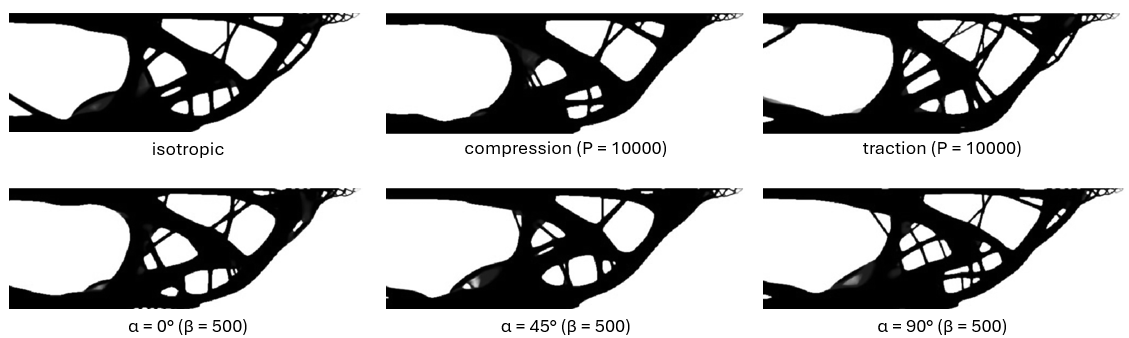}
\caption{\centering Cantilever beam (600x200) with mixed nondeterministic loading conditions (Figure \ref{fig:cantilever_low} top row). Row-wise, from left to right: isotropic (configuration 2 with $\kappa _f=5\cdot10^{-5}$, $\sigma _f=5\cdot10^{-2}$ and $\alpha =9\cdot10^{-3}$), compression and traction (both with $P=10000$) and printing angles of 0$^{\circ}$, 45$^{\circ}$ and 90$^{\circ}$ (with $\beta = 500$). Iteration 100.
}
\label{fig:cantilever_high}
\end{figure}

For criteria effects under uncertain loading, see Figure \ref{fig:cantilever_low} where the distributed loads' position are given by $\mathcal{N}(550,50)$ on the top right corner and $\mathcal{N}(300,50)$ in the bottom middle. Again, criteria reinforce their target mechanical responses (traction/compression) or printing directions (0$^{\circ}$/ 45$^{\circ}$/90$^{\circ}$), although not as clearly due to randomness introduced by non-static loading. If the modulating scalar parameters $P, \beta$ are increased (see Figure \ref{fig:cantilever_high}), their effects become sharper. Since the load distribution is probabilistic, slight differences can be noted on the contour of some topologies, despite nominally being under the same loading conditions. 

Intermediate density pockets, convoluted topologies and feeble twigs are a direct result of uncertainty in load position forcing continuous gradient changes during the TO process and so withdrawing and restoring volume alternatively in the same region. This induces heavy ramification to support every individual load, which can be corrected via filtering. 

\section{Conclusions}

In this article, some of the most common weaknesses of traditional deterministic topology optimization schemes have been addressed: vulnerability to damage, oversizing and unrealistic material layouts due to a poor grasp of uncertainty and material conditions. An easily-implementable solution based on statistics has been proposed, addressing both uncertainty and damage concerns. The examples shown as results prove the robustness of the presented method when compared to the vanilla configuration, providing uncertainty- and damage-proof alternatives to traditional deterministic designs. At the same time, the proposed methodology can be combined with multi-objective optimization (multi-axial, uncertain load in modulus, frequency and position). This upgrade greatly eases the inverse design of flexible and damage-ready topologies, without incurring in an excessive computational cost.\\

By progressively enforcing compliance penalization through cycles and loads, the algorithm can now effectively "learn" how to react to randomly distributed loads, impacts, vibrations, alternating loads and other eventualities by statistically conveying their individual structural implications onto the optimization process. This "learning process" is demonstrably capable of optimizing topologies under a statistically-predictable set of multi-axial loads, as well as adapting its results for important loading changes and/or previously unseen loads, which can be construed as a "validation" phase. These implications, while partial and perhaps reductionist in some aspects, entail interesting implications to be explored in subsequent settings with varied loading, boundary conditions, materials, etc.

Concerning the optimized design's vulnerability to different material conditions commonly ignored by the traditional TO framework (fatigue, mechanical response, manufacturing techniques, elastic/plastic regions, etc.), several criteria have been proposed to reinforce TO designs by penalizing the target variable accordingly (in this case, compliance). This effectively shapes material distribution according to the desired response without incurring in excessive additional computation power either. The proposed criteria's versatility is tested under several boundary conditions, in an individual and combined manner, evaluating its purposes computationally and experimentally with satisfactory results, fully expandable to 3D applications. Future developments shall contemplate more complex Finite Element formulations, including non-linear behaviors (hyperelasticity, plasticity, buckling, etc.) and extension to tridimensional problems with various meshes.



\section*{Funding}

\includegraphics[width=0.15\linewidth]{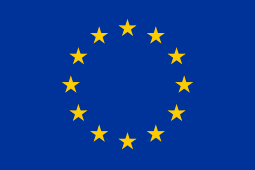}\includegraphics[width=0.33\linewidth]{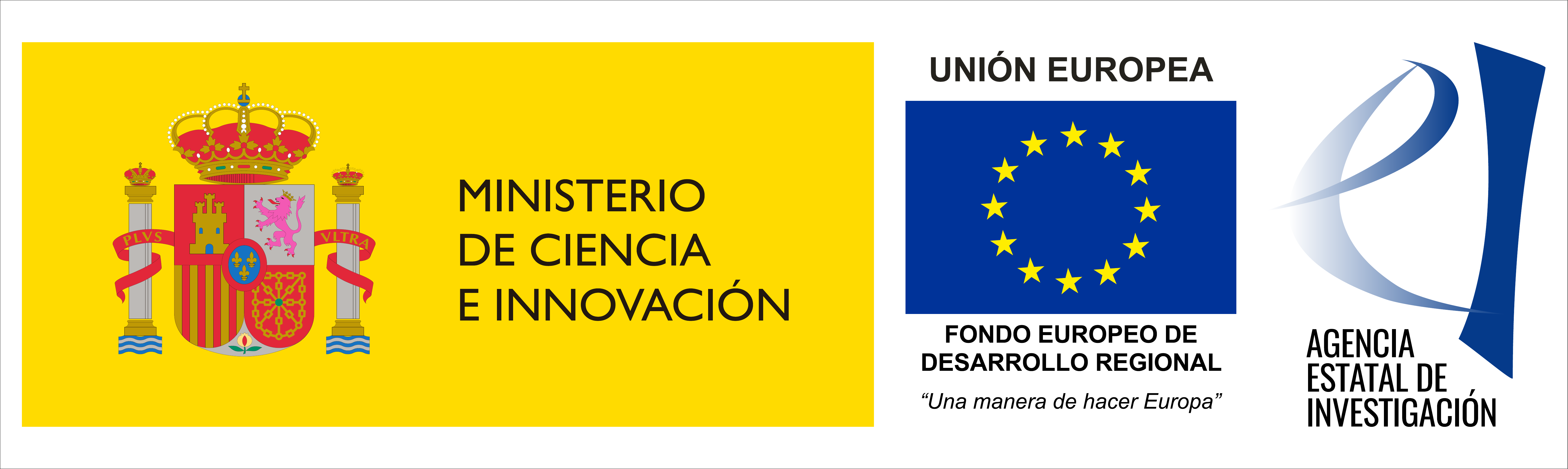}

This project has received funding from the European Union’s Horizon 2020 Marie Sk\l{}odowska-Curie Actions - Innovative European Training Networks under grant agreement No 956401, as well as from the Spanish Ministry of Science and Innovation and the State Research Agency (AEI) PID2021-126051OB-C43.\\

\section*{Conflict of interest}

The authors declare that the research was conducted in the absence of any commercial or financial relationships that could be construed as a potential conflict of interest.

\bibliographystyle{plain}
\bibliography{article.bib}

\end{document}